\documentstyle[11pt]{article}

\newcommand{\nc}{\newcommand}
\nc{\be}{\begin{equation}}
\nc{\ee}{\end{equation}}
\nc{\bea}{\begin{eqnarray}}
\nc{\eea}{\end{eqnarray}}
\nc{\RR} {\rangle }
\nc{\LL} {\langle }
\nc{\half}{\mbox{\small$\frac12$}}
\nc{\beff}{\beta_{\rm eff}}
\nc{\EW}[1]{\langle #1 \rangle}
\nc{\XX}{\phantom{0}}
\nc{\mc}{\multicolumn}

\begin{document}

\begin{titlepage}
\begin{flushright}
HUB-EP-97/22 \\
MS-TPI-97-1  \\
\end{flushright}
\vspace{3ex}
\begin{center}

{\LARGE \bf
\centerline{The Interface Tension of the 3-Dimensional Ising}
\centerline{Model in the Scaling Region}
}

\vspace{1.2 cm}
{\large M. Hasenbusch,}

\vspace{0.3 cm}
{\it Fachbereich Physik, Humboldt Universit\"at zu Berlin,} \\
{\it Invalidenstr.\ 110, 10099 Berlin, Germany} \\[2mm]
email: hasenbus@birke.physik.hu-berlin.de

\vspace{0.8 cm}
{\large K. Pinn,}

\vspace{0.3 cm}
{\it  Institut f\"ur Theoretische Physik I, Universit\"at M\"unster,} \\
{\it  Wilhelm-Klemm-Str.\ 9, D-48149 M\"unster, Germany} \\[2mm]
email: pinn@uni-muenster.de

\vspace{1. cm}
\end{center}
\setcounter{page}{0}
\thispagestyle{empty}
\begin{abstract} \normalsize
Using the Monte Carlo method, we determine  the free energy of the
interface of the 3D Ising model in the scaling region. By integrating
the interface energies over the  inverse temperature $\beta$, we obtain
estimates for the  free energies of interfaces with cross sections up to
96 by 96, and for a range $0.223 \leq \beta \leq 0.23$.  Our data yield a
precise estimation of the interface tensions $\sigma$. We determine the
amplitude $\sigma_0$ in the critical law $\sigma \sim \sigma_0 \,
t^{\mu}$ and estimate the combination $\sigma \xi^2$
which yields the universal constant $R_{-}$ in the
critical limit.
\end{abstract}

\nopagebreak
\vspace{1ex}
\end{titlepage}
\newpage

\section{Introduction}
The numerical determination of interface free energies
and interface tensions in
statistical models is not straightforward. Part of the
difficulties stem from the fact that the interface
free energy is the logarithm of a ratio of partition functions
rather than an expectation value. This prohibits a
straightforward application of the Monte Carlo method.
Various methods have been invented to overcome this
difficulty, each of which has its own merits and disadvantages.
For a number of numerical studies of the 3D Ising interfaces,
all done in the early nineties, see,
e.g., \cite{Klessinger,BergHans,tension,Sanie,direct,ItoPhysA,comparison}.

In this paper, we pick up our methods from~\cite{tension}, focusing
on the critical region of the model.
We present data for the free energies
and interface tensions of
interfaces in the 3D Ising model.
The basic idea,
which is quite old by now, see e.g.~\cite{stauffer},
is to compute the derivative of the interface free energy with
respect to the inverse temperature $\beta$, i.e., the interface energy.
The interface energy is the difference between the energies of
the system with periodic and with antiperiodic
boundary conditions (in one of the lattice directions), respectively.
The interface free energy is then obtained by (numerical)
integration over the energy estimates.

A major advantage of the integration method is that it allows
to study interfaces of quite large cross sections.
This is necessary if one wants to get systematic
errors in the estimates of the interface tension
under control, especially in the scaling region.

In the present paper, we extend our previous results
\cite{tension,comparison} in several ways:
\begin{itemize}
\item
We combine the integration method with
the boundary flip method~\cite{direct}. The latter serves to
provide the initial values for the integration.
\item
We include new efficient algorithms in the simulation
of the systems with antiperiodic and periodic boundary
conditions.
\item
We increase significantly the statistics of the simulations.
\item
We include larger cross sections of the interfaces in
the analysis.
\item
Due to the better precision of our data we can do more
refined fits to theoretical predictions for the quantities
in question.
\end{itemize}

This article is organized as follows:
In Section~\ref{interfaces} we introduce the model, notation
and setup. Furthermore, we discuss the essentials of our
method to compute interface properties and also discuss
finite size effects.
Section~\ref{algorithms} is devoted to a short description
of our Monte Carlo procedures. In section~\ref{freene},
we explain how we determine the interface free energy estimates.
These are the basis for the estimation of the interface tensions
by various fitting procedures. This is discussed in section~\ref{intert}.
Section~\ref{sigma0} reports on our fits to determine the
amplitude $\sigma_0$ and other parameters in the critical law.
It follows section~\ref{Rminus}, where we estimate
the universal amplitude ratio $R_{-}$.
We close with a short summary.

\section{Interfaces in the Ising Model}
\label{interfaces}

Consider the 3D Ising model on
the simple cubic lattice, with the Hamiltonian
\be\label{isiham}
H = - \sum_{<x,y>} s_x s_y  \, , \qquad \quad
s_x = \pm 1 \, .
\ee
The sites of the lattice are labelled by integer coordinates
$x=(x_1,x_2,x_3)$.
The sum in eq.~(\ref{isiham})
is over all (unordered) nearest neighbour pairs of sites
in the lattice. The partition function is
\be
Z = \sum_{\{s\}} \, \exp \left( - \beta H \right) \, .
\ee
Here, the summation is over all possible configurations of
the Ising spins.
The pair interaction is normalized such that $\beta=1/(k_B T)$, where
$k_B$ denotes Boltzmann's constant,
and $T$ is the temperature.

At a critical coupling $\beta_c$ (the estimate of a recent
study~\cite{talapov} is $\beta_c=0.221 6544(6)$) the
model undergoes a second order phase transition.
For $\beta > \beta_c$, the system shows  spontaneous
breaking of the reflection symmetry.

In order to study interfaces separating extended domains of different
magnetization, we consider lattices with extension  $L_1 = L_2 =L$
in the $x_1$- and $x_2$-directions and with extension
$L_3$ in the $x_3$-direction. We
generalize eq.~(\ref{isiham}) to
\be
  H= - \sum_{<x,y>} k_{xy} \, s_x s_y  \, .
\ee
The lattice becomes a torus by regarding the opposite boundary planes
as neighbour planes. In addition to periodic boundary conditions, where
$k_{xy}=1$ for all  links, we consider so called antiperiodic
boundary conditions in $x_3$-direction.
Antiperiodic boundary conditions in $x_3$-direction are imposed  by
$k_{xy}=-1$ for the  links that connect the uppermost
with the lowermost plane. For the other links we keep $k_{xy}=1$.  In the
following we will indicate the type of boundary condition by the subscript $p$
for periodic and $a$ for antiperiodic.
In particular, the Hamiltonian with periodic boundary conditions is denoted
by $H_p$, and the Hamiltonian with antiperiodic boundary conditions by $H_a$.

For sufficiently large $\beta$ and large enough $L$, the imposure of
antiperiodic boundary conditions forces the  system to develop exactly
one interface, a region where the magnetization rapidly changes sign.
This interface is parallel to a (001) lattice plane.

Let us mention that
the Ising (001) interface undergoes a {\em roughening transition} at an
inverse temperature $\beta_R $ that is nearly
twice as large as the bulk transition coupling $\beta_c$.
The presently most accurate estimate  of the inverse roughening
temperature is $\beta_R= 0.40758(1)$~\cite{matching}.
In the region of $\beta_c < \beta < \beta_R$ the interface is
rough. It is smooth (rigid) for $\beta > \beta_R$.

\subsection{Definition of the Interface Free Energy}

We define the interface free energy
as the difference of the free energy
of a system with antiperiodic boundary conditions,
$F_a=-\ln Z_a$, and the free energy of a system
of the same size but periodic boundary conditions, $F_p=-\ln Z_p$, viz.
\be
F_s = F_a - F_p + \ln L_3 \, .
\ee
The term $\ln L_3$ takes care of the possible translations of the interface
along the 3-direction.
For a more detailed discussion of this definition,
see refs.~\cite{tension,direct,comparison,roughbeyond}.

\subsection{Finite Size Behaviour of the Interface Free Energy}
A detailed discussion of
the dependence of $F_s$ on $L_3$ can be found in ref.~\cite{comparison},
section 2. One finds that for
\be
\label{blobi}
 \xi_b <\!\!< L_3 <\!\!< \xi_t \, ,
\ee
$F_s$ is essentially independent of $L_3$.
Here, $\xi_b$ is the bulk correlation length, and $\xi_t$ denotes
the tunneling correlation length.

In ref.~\cite{comparison} we found convergence of the interface energy
within our
numerical accuracy for about $L_3 > 15 \; \xi_b$. The convergence
was found for a large range of interface extensions $L$.
Since $\xi_t = \mbox{ \small $\frac{1}{2}$} \exp(F_s)$,
it is usually no problem to fulfill
the second inequality of eq.~(\ref{blobi}).

The dependence of $F_s$ on $L$ can be discussed in the framework of
effective interface models. It is in a natural way related to
the question of how to define and determine the interface tension $\sigma$.
Note that $\sigma$ can unambiguously be defined
through
\be
\label{origdef}
\sigma_{\infty} = \lim_{L \rightarrow \infty} \frac{F_s}{L^2} \, .
\ee
In order to extract reliable estimates of $\sigma_{\infty}$ from
finite $L$ data (cf.~section~\ref{intert} below),
one profits very much from a more detailed
information on the finite $L$ behaviour of $F_s$.

It was observed already in~\cite{tension} that
for sufficiently large $L$, $F_s$ follows
with good precision
\be
\label{sigmadef}
F_s \simeq C_s + \sigma \, L^2  \, ,
\ee
where $C_s$ is a constant. This type of behaviour is suggested
by the 1-loop approximation of the capillary
wave model~\cite{gelfandfisher,privman,roughbeyond}
and also by the 1-loop semiclassical expansion
in the field theoretic framework~\cite{gernot}.

In \cite{roughbeyond} it was derived from the 2-loop-expansion
of the capillary wave model that the free energy should behave as
\be
\label{2loop}
F_s \simeq C_s + \sigma \, L^2 - \frac{1}{4 \sigma L^2} \, .
\ee

\subsection{How to Compute $F_s$}

In this subsection we outline
our strategy to compute the interface free energy.
The boundary flip algorithm allows for a direct Monte Carlo measurement
of $F_s$. However, the method works well only for
$F_s < 10$. Therefore we combined
the boundary flip method with the integration method
first used in ref.~\cite{stauffer}.
Taking the derivative of $F_s$ with respect to $\beta$,
we obtain the interface energy $E_s$,
\be
E_s = E_a - E_s \; ,
\ee
where $E_a$ and $E_p$ are the expectation values of the corresponding
Hamiltonians, $H_a$ and $H_p$.
$E_a$ and $E_p$ can be computed by  Monte Carlo
simulations of systems with antiperiodic and periodic boundary conditions,
respectively.
The interface free energy in a range of inverse temperatures
is then obtained by numerically performing the integration
\be
F_s(\beta) = F_s(\beta_0) + \int_{\beta_0}^{\beta} d\beta'
\,  E_s(\beta') \; .
\ee
Note that one could in principle choose any $\beta_0$ as
the starting point of the integration. We chose $\beta_0$ close
to criticality, so that the condition $F_s < 10$ is fulfilled
also for the larger interface cross sections (cf.~table~\ref{start}
and section~\ref{freene} below).

\section{The Monte Carlo Algorithms}

\label{algorithms}

For the simulations we used four types of algorithms:

\begin{itemize}
\item
boundary flip algorithm
\item
single cluster algorithm
\item
surface cluster algorithm
\item
demon algorithm
\end{itemize}

The boundary flip algorithm was used to determine the interface free energy
at the smallest inverse temperature $\beta$ considered for a given lattice
size. For the measurement of the energy with periodic and antiperiodic boundary
conditions we simulated the system with
the single cluster algorithm~\cite{ulli}
(combined with a surface cluster algorithm~\cite{surface}
in case of antiperiodic boundary conditions)
for a $\beta$-range
closest to the critical value,
while the demon algorithm was used for the remaining $\beta$-values up to
$\beta=0.23$.

The boundary flip algorithm~\cite{direct}
allows to simulate an ensemble that contains
periodic as well as  antiperiodic boundary conditions. The partition
function
of this system is given by $Z=Z_p+Z_a$. The ratio $Z_a/Z_p$ is given by
$<\delta_{b,a}>/<\delta_{b,p}>$, where $\delta_{b,a}$ is equal to one if the
boundary conditions are antiperiodic, and zero else.
$\delta_{b,p}$ is defined analogously. The boundary flip algorithm is a
modification of the cluster algorithm.
For a detailed discussion of the algorithm see ref.~\cite{finite}.

The single cluster algorithm was introduced in~\cite{ulli}
as an improvement of the multi-cluster algorithm
of Swendsen and Wang~\cite{sw}. For a detailed
discussion of its critical dynamical behaviour see~\cite{ulli2}.

In case of the antiperiodic boundary conditions, and thus the
existence of an interface,
a straightforward application of the bulk cluster
algorithm is not appropriate.
The interface is correlated on all length scales,
whereas the bulk correlation
length is finite. Furthermore, competing interactions
are induced by the interface, and cluster algorithms
become inefficient for frustrated systems.
We therefore combined the single cluster algorithm with
the interface cluster
proposed in ref.~\cite{surface},
and a slight modification of it which is more suitable for
simulations close to the bulk critical temperature.

For the larger $\beta$-values
we used a demon algorithm~\cite{creutz} combined with canonical updates of the
demons~\cite{kari}. The algorithm is a local algorithm.
Its advantage is that compared to standard algorithms one needs much
less random numbers.

The algorithm
can easily be implemented in multi-spin coding fashion. For each spin
(and demon) only one bit is used, and the operations are done simultaneously
on each bit of a given word (in our case 64 bits per word).
For a detailed discussion of the algorithm see ref.~\cite{comparison}.

In order to give an idea of the CPU-times required
we give some typical update times
for a DEC Alpha 250 4/266 workstation,
where most of our simulations were performed.
One boundary flip update for the $96^3$ lattice at $\beta=0.2219$ takes
0.515 sec. The update of a single site with the single cluster algorithm
takes $1.9 \times 10^{-6}$ sec. This time should be compared with
the performance of the demon program in multispin coding:
Here the update of a single spin takes
$46 \times 10^{-9}$ seconds on an HP 735 and
$21 \times 10^{-9}$ seconds on a DEC Alpha 250 workstation, measured on
a $120^3$ lattice.
For $\beta=0.22311$ the integrated
autocorrelation time of the magnetization was $\tau_{int} = 81(2)$ in
units of sweeps.

\section{Determination of Interface Free Energies}
\label{freene}

In order to obtain start values for the integration over $\beta$
we simulated the 3D Ising model with fluctuating boundary conditions.

The $\beta$-values were chosen to be close to $\beta_c$, since
here the interface free energies are small also on the larger lattices,
and the boundary flip algorithm works still very well.

We chose eight different lattice volumes: The spatial extension
(denoted by $L$ in this paper) was 32, 48, 64, and 96. For the
temporal extension (denoted by $L_3$) we used always two different
values, namely $L_3=L/2$ and $L_3=L$. The data for the smaller
extension in $x_3$-direction were taken in order to have a measure
for the finite $L_3$ effects.

Our estimates together with a specification of the number of
boundary flip updates to obtain them are given in
table~\ref{start}.

The next task then was to perform the numerical integration
over the interface energies. Using the various algorithms
described before, we obtained a large number of estimates
for the interface energy, distributed on a fine grid of
$\beta$-values, typically spaced by $\Delta \beta = 0.00005$
or $0.0001$, see again table~\ref{start} for details.
The numerical integration was done using the trapezoidal rule.
For a check we also used interpolation with splines, yielding
consistent results. Part of the integration was also done
employing the ``finite step $\Delta \beta$ method'' described
in~\cite{tension}.

After the laborious procedure we ended up
(for each of the cross sections 32,48,64,96) with 71 estimates
for the interface free energy, in the range $0.223 \leq \beta \leq 0.23$,
spaced by $\Delta \beta = 0.0001$.
For a selection of a few estimates, see our
table~\ref{freeestimates}).
We decided not to quote our
estimates for $\beta < 0.223$, since they seem to suffer from
visible finite $L_3$ effects. Also, the data with $L_3=L/2$, that
were taken for $L=32$ and $L=96$ were after some inspection
discarded from the further analysis.

\section{Determination of the Interface Tension $\sigma$}
\label{intert}

In order to obtain estimates for the interface tensions,
we employed eqs.~(\ref{sigmadef})~and~(\ref{2loop}).

We did five types of fits, to be labelled by {\bf fit1}, {\bf fit2},
{\bf 64vs48}, {\bf 96vs64}, and {\bf fit3}. The various fit
types are defined in the following table.

\vspace{1cm}
\begin{tabular}{ll}
 {\bf fit1}   &   Fit the free energy data for $L$=32, 48, 64, and 96
                  with eq.~(\ref{sigmadef}) \\[5mm]
 {\bf fit2}   &   Fit the free energy data for $L$= 48, 64, and 96
                  with eq.~(\ref{sigmadef}) \\[5mm]
 {\bf 64vs48} &   Compute $C_s$ and $\sigma$ of eq.~(\ref{sigmadef})
                  from the $L=64$ and $48$ data alone \\[5mm]
 {\bf 96vs64} &   Compute $C_s$ and $\sigma$ of eq.~(\ref{sigmadef})
                  from the $L=96$ and $64$ data alone \\[5mm]
 {\bf fit3}   &   Fit the free energy data for $L$=32, 48, 64, and 96
                  with eq.~(\ref{2loop}) \\[5mm]
\end{tabular}

\vspace{1cm}

These fits were applied to our free energy estimates
($L=32, 48, 64, 96$) for the 71 different $\beta$-values specified
in the previous subsection.
Our fit results for
$\sigma$ for a selection of $\beta$-values are given in
table~\ref{sigma_estimates}.
With a few exceptions, the $\sigma$-estimates of the various
fit types seem fairly consistent with each other, indicating that,
for the range of $\beta$-values chosen, the effects from
finite interface cross sections seem under control.
Being cautious in estimating systematic errors, we have
a tendency to announce the {\bf 96vs64} fit results as our
final estimates for $\sigma$.
The fits with the 2-loop approximation eq.~(\ref{2loop})
is, however, also very interesting.
For $\beta=0.223$ we did a more detailed analysis,
comparing {\bf fit1} and {\bf fit3},
with further cross sections included. The results are
summarized and explained in table~\ref{fits_sigma}.
It seems that the fits with the 2-loop formula are
more stable with respect to inclusion of smaller
$L$-values, thus supporting the claims done in
ref.~\cite{roughbeyond}.

In~\cite{hoppemuenster} it is suggested that the interface
free energy should contain an extra term with
a logarithmic $L$-dependence,
\be
F_s = C_s + \sigma L^2 - \kappa \ln L + ...
\ee
with $\kappa \approx 1.65$. We note that our data
are incompatible with such a big logarithmic correction.

It is interesting to look at the combination
\be
G= C_s + \frac12 \, \ln \sigma \, .
\ee
This quantity has a scaling limit that can be compared with
results of the semiclassical calculation of M\"unster~\cite{gernot}.
This calculation yields
\be
G = \ln 2 - \ln S \, ,
\ee
with
\be
S = 4 \frac{ \Gamma\left(\frac34 \right) }
           { \Gamma\left(\frac14 \right) }
      \left( 1 - \frac{u_R}{4 \pi}
      \left( \frac {39}{32} - \frac{15}{16} \ln 3 \right) \right)^{-1/2}
\, .
\ee
Assuming that $u_R = 14.3(1)$ in the critical limit~\cite{michelemartin},
the semiclassical prediction (to this order) is
\be
G \approx 0.29 \, .
\ee
Our numerical estimates for this quantity
are given in table~\ref{constln}.
Note that within the statistical errors there is a
nice agreement of our results with the theoretical prediction.

\section{Determination of $\sigma_0$}
\label{sigma0}

We fitted our estimates for $\sigma$ with the ansatz
\be
\label{critlaw}
\sigma(\beta) = \sigma_0 \;t^\mu \;
\left( 1 \, + \, a_{\theta} \, t^\theta \, + \, a_1 \, t \right) \, .
\ee
In this ansatz, the leading corrections to scaling are included.
Fits according to eq.~(\ref{critlaw}) were recently applied
by Zinn and Fisher~\cite{ZinnFisher} to our data
published in~\cite{tension}.

We compared two different definitions of $t$, namely
\bea
t_1 &=& \beta/\beta_c - 1 \\
t_2 &=& 1 - \beta_c/\beta  \, .
\eea
Agreement of fits with the two different definitions
could be interpreted as a signal that effects from
finite $t$ (and/or higher order corrections) are under control.
Note also that the coefficient $a_1$ should jump by $\mu$ when
changing the definition of $t$.

In our fits, we always fixed the values of
$\beta_c$, $\mu=2\nu$, and $\theta$.
We consider the following choice of these parameters as reasonable:
$\beta_c = 0.2216544(6)$ \cite{talapov}, $\mu=2 \times 0.631$,
$\theta =0.51(3)$ \cite{talapov}.

To compare the different definitions of $\sigma$, cf.~section~\ref{intert},
we did the fits for all the definitions.
Always all 71 $\beta$-values were included in the fit. All the
fits had a very good $\chi^2/$dof (around one or smaller).
Discarding
the $\beta$-values farer away from criticality did not improve
the quality of the fits.

Note that the $\sigma$-estimates for the various $\beta$-values
are not completely statistically independent, since they stem
{}from an integration procedure over the energy data.
In order to take into account the cross-correlations properly,
we redid all fits on a set of 50
suitably generated baby data sets for the $\sigma$'s.
The statistical estimate for the fit parameters was then
obtained from the variance over the 50 babies.

In table~\ref{sigmanull} we give our results for the above quoted
choice of the fixed parameters $\beta_c$, $\nu$, and $\theta$.
One immediately recognizes that the estimates for the
two different $t$-definitions agree nicely. However, the
variation with the
fit types seems a bit stronger. If we disregard the
{\bf 64vs48} which does not include the $L=96$ data,
the results for $\sigma_0$ scatter from 1.54 to 1.57.

Of course, one has to check also the dependence of the fit results
on the parameters fixed in the fit procedure. We found that
the dependence on the choice of $\beta_c$ and $\theta$ is
quite weak, whereas the dependence on
$\mu = 2 \nu$ turns out to be non-negligible. We thus quote
the results for one type of fit ({\bf fit1}) for a range
of ``reasonable'' $\nu$-values in table~\ref{deponnu}.
The estimate for $\sigma_0$ moves more or less from 1.51 to 1.59,
whereas the relative variation  of the other fit parameters is
even stronger.
Doing the same comparison with type {\bf 96vs64} yielded the
same range of estimates for $\sigma_0$.

In summary, taking into account the systematic dependencies
on the fit type and on the dependence on
the input of the exponent $\nu$ we quote as our
final estimate
\be
\sigma_0 = 1.55(5) \, .
\ee
It is interesting to compare our result with those
of the literature. We compiled a few of them in
table~\ref{sig0comp}. A fair agreement with most
of the more recent estimates is found.

\section{The Universal Amplitude Ratio $R_{-}$}
\label{Rminus}

Using the most recent numerical results for the second moment correlation
length $\xi_{\rm 2nd}$ \cite{michelemartin}
we computed estimates for the universal amplitude ratio
\be
R_{-} = \lim_{\beta \searrow \beta_c} \sigma(\beta) \,
        \xi_{\rm 2nd}(\beta)^2 \, .
\ee
In order to extract the limiting value from our data we fitted according to
\be
\sigma(\beta) \, \xi_{\rm 2nd}(\beta)^2 = R_{-} + c \; \xi^{-\omega} \, ,
\ee
where we take the numerical value $\omega=0.81(5)$
from the literature~\cite{talapov}.
Including all $\beta$-values of table~\ref{rminus}, we obtain
$R_{-} =0.1040(8)$ and  $c=-0.023(1)$. The corresponding fit
has a $\chi^2/$dof of $0.55$.
These results remain stable
when the largest $\beta$-values are discarded.
The error is dominated by the uncertainty of $\omega$.

We again would like to compare our estimate with others in
the literature. The perhaps most interesting comparison
is with the semiclassical expansion of M\"unster~\cite{gernot}.
It was recently extended to 2-loop~\cite{hoppemuenster}:
\be
R_{-} = \frac{2}{u_R^*} \left(1 + \sigma_{1l} \frac{u_R^*}{4 \pi}
+ \sigma_{2l} \left(\frac{u_R^*}{4 \pi}\right)^2 + \dots \right) \, .
\ee
with $\sigma_{1l}= -.2002602$ and $\sigma_{2l}=-0.0076(8)$.
Plugging in the most recent result for
$u_R^* =14.3(1)$ \cite{michelemartin} one obtains for
1-loop $R_{-,1l} =0.1080(10)$, and on 2-loop level
$R_{-,2l} =0.1066(10)$.
In both cases the error is determined by the uncertainty of
$u_R^*$.

It might be also interesting to compare with a few results obtained
by numerical studies, see table~\ref{rmincomp}.

\section*{Summary}

We have presented a numerical study of the 3D Ising interface
tension in the scaling region, using the method of
``integration of the interface energy over $\beta$''.
Based on our results for the interface tension we
estimated the amplitude $\sigma_0$ and the
universal amplitude ratio $R_{-}$. Our results are
$\sigma_0 = 1.55(5)$ and  $R_{-} = 0.1040(8)$, respectively.

\listoftables
\newpage

\begin{table}
\small
\begin{center}
\begin{tabular}{|c|c|c|c|c|l||l|c|}
\hline
\mc{1}{|c}{$\beta_0$} &
\mc{1}{|c}{ $L$}  &
\mc{1}{|c}{$L_3$}  &
\mc{1}{|c}{$stat$ } &
\mc{1}{|c}{ $<$boundary$>$ } &
\mc{1}{|c||}{$F_s$} &
$\beta$-range &
$\Delta \beta/10^{-4}$\\
\hline
0.2219 &96 & 48 &  201,500 &  0.9056(10)  &  6.876(15) & 0.2220-0.23 & 2/0.5 \\
0.2219 &96 & 96 &  201,000 &  0.8324(13)  &  6.956(8)  & 0.2220-0.23 & 2/0.5 \\
\hline
0.2220 &64 & 32 &  209,000 &  0.8560(12)  &  6.022(9)  &                 &   \\
0.2220 &64 & 64 &  213,300 &  0.7063(17)  &  5.918(7)  & 0.2221-0.23 & 1/0.5 \\
\hline
0.2225 &48 & 24 &  200,000 &  0.9306(10)  &  6.503(15) &                 &   \\
0.2225 &48 & 48 &  200,000 &  0.8812(12)  &  6.633(11) & 0.2227-0.23 & 0.5   \\
\hline
0.2230 &32 & 16 &  200,000 &  0.8999(11)  &  5.716(12) & 0.2256-0.23 & 0.5   \\
0.2230 &32 & 32 &  225,000 &  0.8112(14)  &  5.727(8)  & 0.2230-0.23 & 0.5   \\
\hline
\end{tabular}
\parbox[t]{.85\textwidth}
 {
 \caption[Free energy estimates from boundary flip simulations]
 {\label{start}
  Interface free energy $F_s$ for $\beta$ close to $\beta_c$ obtained
  from boundary
  flip simulations as described in ref.~\cite{direct}.
  $stat$ denotes the number
  of boundary flip updates,  $<$boundary$>$ is the
  expectation value of the boundary observable, which takes the value $-1$
  for antiperiodic boundary and
  $+1$ for periodic boundary conditions.
  %$-\ln(Z_a/Z_p) + \ln L_3 $.
  In the r.h.s.\ of the table we specify the range of $\beta$-values
  where Monte Carlo estimates for the interface energy were obtained.
  The $\beta$-values were spaced by increments $\Delta \beta$.
 }
 }
\end{center}
\end{table}

\begin{table}
\begin{center}
\begin{tabular}{|r|r|r|r|r|}
\hline
\mc{1}{|c} {$\beta$} &
\mc{1}{|c} {$L=32$}  &
\mc{1}{|c} {$L=48$}  &
\mc{1}{|c} {$L=64$} &
\mc{1}{|c|}{$L=96$} \\
\hline
 0.2230 &   5.727(08) & 8.796(12) & 13.088(27) &  25.377(64) \\
 0.2235 &   6.717(08) &11.295(12) & 17.611(34) &  35.879(71) \\
 0.2240 &   7.844(08) &13.997(13) & 22.535(37) &  47.104(80) \\
 0.2247 &   9.583(09) &18.051(14) & 29.848(37) &  63.725(85) \\
 0.2255 &  11.718(09) &22.990(14) & 38.748(38) &  83.919(85) \\
 0.2265 &  14.565(09) &29.544(15) & 50.517(38) & 110.563(86) \\
 0.2275 &  17.567(10) &36.443(16) & 62.870(39) & 138.482(87) \\
 0.2285 &  20.706(10) &43.620(17) & 75.681(40) & 167.441(88) \\
 0.2300 &  25.610(11) &54.805(18) & 95.639(41) & 212.535(89) \\
\hline
\end{tabular}
\parbox[t]{.85\textwidth}
 {
 \caption[Selection of free energy estimates]
 {\label{freeestimates}
  A selection of our interface free energy estimates
  computed in the range $0.223 \leq \beta \leq 0.23$.
 }
 }
\end{center}
\end{table}

\begin{table}
\begin{center}
\begin{tabular}{|l|l|l|l|l|l|}
\hline
fit &
$\beta=0.2230$ &
$\beta=0.2240$ &
$\beta=0.2255$ &
$\beta=0.2275$ &
$\beta=0.2300$ \\
\hline
{\bf fit1}  &0.002398(5) & 0.004793(6) & 0.008808(6) & 0.014753(9)
 & 0.022810(8)  \\
{\bf fit2}  &0.002398(6) & 0.004784(9) & 0.008810(9) & 0.014759(11)
& 0.022813(11) \\
{\bf 64vs48}&0.002395(16)& 0.004765(22)& 0.008794(22)& 0.014747(24)
 & 0.022786(25) \\
{\bf 96vs64}&0.002400(14)& 0.004799(17)& 0.008822(18)& 0.014768(19)
 & 0.022831(19) \\
{\bf fit3}  &0.002376(5) & 0.004782(6) & 0.008801(6) & 0.014750(9)
 & 0.022808(8)  \\
\hline
\end{tabular}
\parbox[t]{.85\textwidth}
 {
 \caption[Selection of interface tension estimates]
 {\label{sigma_estimates}
  Estimates for the interface tension $\sigma$ as obtained
  from fitting the interface energies in various ways. The different
  types of fits are explained in the text.
 }
 }
\end{center}
\end{table}

\begin{table}
\begin{center}
\begin{tabular}{|r|r|r|l|l||l|l|r|}
\hline
\mc{1}{|c|}{$L$} &
\mc{1}{c|}{$F_s$}  &
\mc{1}{c|}{$\chi^2/$dof} &
\mc{1}{c|}{$C_s$} &
\mc{1}{c||}{$\sigma$}  &
\mc{1}{c|}{$\sigma$}  &
\mc{1}{c|}{$C_s$} &
\mc{1}{c|}{$\chi^2/$dof} \\
\hline
 8 &   2.669(09) & 638.6& 2.898(5) & 0.002554(5) & 0.002341(5)
&3.681(6)&715.7\\
12 &   3.218(08) & 240.7& 3.053(5) & 0.002486(5) & 0.002346(5) &3.531(6)&
42.9\\
18 &   3.912(11) &  35.2& 3.222(7) & 0.002420(5) & 0.002367(5) &3.458(8)& 2.0
\\
24 &   4.641(11) &   7.1& 3.288(9) & 0.002399(6) & 0.002369(6) &3.453(9)& 2.2
\\
30 &   5.459(13) &   3.5& 3.325(12)& 0.002390(6) & 0.002371(6) &3.446(12)&3.1
\\
36 &   6.457(17) &   2.5& 3.362(21)& 0.002382(7) & 0.002369(7) &3.455(20)&5.8
\\
64 &  13.088(27) &      & 3.257(50)& 0.002400(14)& 0.002397(14)&3.293(50)&
\\
96 &  25.377(64) &      &          &             &             &         &
\\
\hline
\end{tabular}
\parbox[t]{.85\textwidth}
 {
 \caption[Fits for the interface tension]
 {\label{fits_sigma}
  A comparison of two different fits to interface free energy data a
  $\beta=0.223$. The second column gives the free energy estimates
  for the lattice cross sections ranging from $L=8$ to
  $L=96$. The data for $L=64$ and $L=96$ are taken from the
  present study, the estimates for the smaller lattices are taken
  from ref.~\cite{direct}. Columns 3, 4 and 5 give the results
  of a fit with eq.~(\ref{sigmadef}). The first
  line gives the fit results when all data, starting from
  the $L=8$ value, are included. The second lines gives
  the result when the $L=8$ result is excluded from the
  data, the third line is based on discarding
  $L=8$ and $L=12$, and so on. In the last estimate only the two largest
  lattice sizes enter.   In the right hand part of the table,
  we give for comparison the results for a fit with eq.~(\ref{2loop}).
 }
 }
\end{center}
\end{table}

\begin{table}
\begin{center}
\begin{tabular}{|l|l|l|l|l|l|}
\hline
fit &
$\beta=0.2230$ &
$\beta=0.2240$ &
$\beta=0.2255$ &
$\beta=0.2275$ &
$\beta=0.2300$ \\
\hline
{\bf fit1}  & 0.255(10) & 0.269(11) & 0.332(12) & 0.349(16) & 0.361(17) \\
{\bf fit2}  & 0.253(21) & 0.300(26) & 0.323(29) & 0.326(34) & 0.348(32) \\
{\bf 64vs48}& 0.260(44) & 0.346(56) & 0.363(58) & 0.356(63) & 0.415(67) \\
{\bf 96vs64}& 0.241(70) & 0.209(92) & 0.247(96) & 0.273(99) & 0.231(103)\\
{\bf fit3 } & 0.368(10) & 0.328(11) & 0.364(12) & 0.368(16) & 0.373(17) \\
\hline
\end{tabular}
\parbox[t]{.85\textwidth}
 {
 \caption[Results for G = $C_s + \frac12 \ln \sigma$]
 {\label{constln}
  Estimates for the sums $G = C_s + \frac12 \ln \sigma$.
  The different types of fits are explained in the text.
 }
 }
\end{center}
\end{table}

\begin{table}
\begin{center}
\begin{tabular}{|l|l|l|l|r|}
\hline
\mc{1}{|c}{fit} &
\mc{1}{|c}{$t$-def} &
\mc{1}{|c}{$\sigma_0$} &
\mc{1}{|c}{$a_{\theta}$}  &
\mc{1}{|c|}{$a_1$} \\
\hline
{\bf fit1}   & $t_1$ & 1.5677(73) & -0.563(47)  &  0.49(13) \\
             & $t_2$ & 1.5682(75) & -0.561(49)  &  1.64(13) \\
\hline
{\bf fit2}   & $t_1$ & 1.549(11)  & -0.409(72)  &  0.01(20) \\
             & $t_2$ & 1.549(11)  & -0.397(76)  &  1.13(21) \\
\hline
{\bf fit3}   & $t_1$ & 1.5428(73) &  -0.376(49) & -0.06(14) \\
             & $t_2$ & 1.5421(73) &  -0.362(51) &  1.06(14) \\
\hline
{\bf 96vs64} & $t_1$ & 1.571(20) & -0.57(13) &   0.47(36) \\
             & $t_2$ & 1.571(21) & -0.56(13) &   1.61(37) \\
\hline
{\bf 64vs48} & $t_1$ & 1.519(23) & -0.18(16) &  -0.68(44) \\
             & $t_2$ & 1.517(23) & -0.15(16) &   0.42(45) \\
\hline
\end{tabular}
\parbox[t]{.85\textwidth}
 {
 \caption[Fit results for $\sigma_0$]
 {\label{sigmanull}
Fit results for the coefficients $\sigma_0$, $a_{\theta}$, and
$a_1$ in the critical law eq.~(\ref{critlaw}).
In these fits, we fixed the following parameters:
$\beta_c= 0.2216544$, $\mu=1.262$, and $\theta=0.51$.
 }
 }
\end{center}
\end{table}

\begin{table}
\begin{center}
\begin{tabular}{|l|l|l|l|}
\hline
\mc{1}{|c}{$\nu$} &
\mc{1}{|c}{$\sigma_0$} &
\mc{1}{|c}{$a_{\theta}$}  &
\mc{1}{|c|}{$a_1$} \\
\hline
 0.628&  1.5045(70) & -0.394(48) & 0.17(13) \\
 0.629&  1.5253(71) & -0.451(48) & 0.28(13) \\
 0.630&  1.5464(72) & -0.507(47) & 0.38(13) \\
 0.631&  1.5677(73) & -0.563(47) & 0.49(13) \\
 0.632&  1.5892(73) & -0.618(47) & 0.59(13) \\
\hline
\end{tabular}
\parbox[t]{.85\textwidth}
 {
 \caption[Dependence of $\sigma_0$ on $\nu$]
 {\label{deponnu}
Checking the dependence of the fit results for the coefficients
$\sigma_0$, $a_{\theta}$, and $a_1$ on the
variation of $\nu = \mu/2$. The two other parameters,
$\beta_c$ and $\theta$, are fixed to the values
0.2216544 and 0.51, respectively. The fit type is {\bf fit1}
here, and the $t$-type is $t_1$.
 }
 }
\end{center}
\end{table}

\begin{table}
\begin{center}
\begin{tabular}{|c|l|c|l|}
\hline
\mc{1}{|c}{year} &
\mc{1}{|c}{authors(s)} &
\mc{1}{|c}{Ref.} &
\mc{1}{|c|}{$\sigma_0$} \\
\hline
1982 & Binder                   & \cite{BinderS}    & 1.05(5)   \\
1984 & Mon and Jasnow           & \cite{MonJasnow}  & 1.2(1)    \\
1988 & Mon                      & \cite{Mon}        & 1.58(5)   \\
1992 & Klessinger and M\"unster & \cite{Klessinger} & 1.29-1.64 \\
1993 & Berg et al.              & \cite{BergHans}   & 1.52(5)   \\
1993 & Ito                      & \cite{ItoPhysA}   & 1.42(4)   \\
1993 & Hasenbusch and Pinn      & \cite{tension}    & 1.22-1.49 \\
1993 & Hasenbusch               & \cite{direct}     & 1.5(1)    \\
1993 & Gausterer et al.         & \cite{Sanie}      & 1.92(15)  \\
1994 & Caselle et al.           & \cite{roughbeyond}  & 1.32-1.55 \\
1996 & Zinn and Fisher          & \cite{ZinnFisher}     & 1.50(1)   \\
1997 & Hasenbusch and Pinn      & this work         & 1.55(5)   \\
\hline
\end{tabular}
\parbox[t]{.85\textwidth}
 {
 \caption[Comparison of $\sigma_0$ estimates]
 {\label{sig0comp}
  Comparison of a number of estimates for $\sigma_0$ taken
  from the literature. The estimate by Zinn and Fisher is
  based on data from~\cite{tension}.
 }
 }
\end{center}
\end{table}

\begin{table}
\begin{center}
\begin{tabular}{|l|l|l|l|}
\hline
\mc{1}{|c}{$\beta$} &
\mc{1}{|c}{$\xi_{\rm 2nd}$} &
\mc{1}{|c}{$\sigma$}  &
\mc{1}{|c|}{$\sigma \, \xi_{\rm 2nd}^2$} \\
\hline
0.2391  & 1.2335(15) & 0.05555(10)  & 0.0845(3) \\
0.23142 & 1.8045(21) & 0.02760(11)  & 0.0899(4) \\
0.2275  & 2.5114(31) & 0.014768(19) & 0.0931(3) \\
0.2260  & 3.0340(32) & 0.010257(18) & 0.0944(3) \\
0.2240  & 4.509(6)   & 0.004799(17) & 0.0976(4) \\
0.22311 & 6.093(9)   & 0.002649(14) & 0.0983(6) \\
\hline
\end{tabular}
\parbox[t]{.85\textwidth}
 {
 \caption[Results for $\sigma \xi_{\rm 2nd}^2$]
 {\label{rminus}
Results for the combination $\sigma \xi_{\rm 2nd}^2$.   In the second
column we give the results for the second moment correlation length
obtained in  ref.~\cite{michelemartin}. $\sigma$ is our present estimate
for the interface tension. In the last column we give $\sigma \xi_{\rm
2nd}^2$ obtained from  the numerical results for $\xi_{\rm 2nd}$ and
$\sigma$.  The estimates are used to determine the universal amplitude
ratio $R_{-}= \lim_{\beta \searrow \beta_c} \xi_{\rm 2nd}^2 \, \sigma$.
 }
 }
\end{center}
\end{table}

\begin{table}
\begin{center}
\begin{tabular}{|c|l|c|l|}
\hline
\mc{1}{|c}{year} &
\mc{1}{|c}{authors(s)} &
\mc{1}{|c}{Ref.} &
\mc{1}{|c|}{$R_{-}$} \\
\hline
1992 & Klessinger and M\"unster & \cite{Klessinger}   & 0.090(3)   \\
1993 & Hasenbusch and Pinn      & \cite{tension}      & 0.090(5)   \\
1996 & Zinn and Fisher          & \cite{ZinnFisher}   & 0.096(2)   \\
1996 & Agostini et al.          & \cite{Ago}          & 0.1056(19) \\
1997 & Hasenbusch and Pinn      & this work           & 0.1040(8)  \\
\hline
\end{tabular}
\parbox[t]{.85\textwidth}
 {
 \caption[Comparison of $R_{-}$ estimates]
 {\label{rmincomp}
  Comparison of a number of estimates for $R_{-}$ taken
  from the literature. The estimate of Zinn and Fisher
  is based on data of~\cite{tension}. Agostini et
  al.\ used the true instead of
  the second moment correlation length.
 }
 }
\end{center}
\end{table}

\end{document}